\documentclass[aps,prb,10pt,twocolumn,showpacs,superscriptaddress]{revtex4-1}
\usepackage{amssymb,graphicx,color,amsmath,xfrac,bm,stmaryrd,trimclip}
\usepackage[mathletters]{ucs}
\usepackage[utf8x]{inputenc}
\definecolor{lightblue}{rgb}{0.17,0.39,1}
\usepackage[bookmarks, colorlinks=true, breaklinks]{hyperref}
\hypersetup{linkcolor=lightblue, citecolor=lightblue, filecolor=black, urlcolor=lightblue}

\makeatletter
\DeclareRobustCommand{\shortto}{\mathrel{\mathpalette\short@to\relax}}
\newcommand{\short@to}[2]{{\mkern2mu	 \clipbox{0 0 {0.5\width}  0 }{$\m@th#1\vphantom{+}{\shortleftarrow}$}}}
\makeatother

\newcommand{\from}{\!\raisebox{-.05ex}{$\shortto$}}

\renewcommand{\Im}{{\mathrm{Im}}}

\newcommand{\para}[1]{\left(#1\right)}
\newcommand{\brac}[1]{\left[#1\right]}

\newcommand{\av}[1]{\left\langle#1\right\rangle}
\newcommand{\matr}[1]{\brac{\begin{matrix}#1\end{matrix} }}

\newcommand{\fra}[2]{ {#1}/{#2}}
\newcommand{\frap}[2]{\para{ {#1}/{#2}}}

\newcommand{\mcal}{γ}  % magnetocaloric coefficient
\newcommand{\ga}{υ} 
\newcommand{\M}{{\text{M}}}
\renewcommand{\L}{{\text{L}}} 
\renewcommand{\S}{{\text{S}}}
\newcommand{\C}{{\text{C}}}
\newcommand{\B}{{\text{B}}}
\newcommand{\N}{{\text{N}}}
\newcommand{\A}{\text{A}}

\newcommand{\observed}{{\text{observed}}}
\newcommand{\model}{{\text{model}}}
\newcommand{\eff}{{\text{eff}}}
\newcommand{\NfromS}{\N\from\S}
\newcommand{\SfromC}{\S\from\C}
\newcommand{\LfromC}{\L\from\C}
\newcommand{\MfromC}{\M\from\C}
\newcommand{\CfromB}{\C\from\B}

\begin{document}
\title{Calorimetric measurement of nuclear spin-lattice relaxation rate in metals}

\author{A. Khansili}
\affiliation{Department of Physics, Stockholm University, SE-106 91 Stockholm, Sweden}

\author{A. Bangura}
\affiliation{National High Magnetic Field Laboratory, Tallahassee FL 32310}

\author{R. D. McDonald}
\affiliation{Los Alamos National Laboratory, Los Alamos NM 87545}

\author{B. J. Ramshaw}
\affiliation{Laboratory of Atomic and Solid State Physics, Cornell University, Ithaca NY 14853 USA}

\author{A. Rydh}
\affiliation{Department of Physics, Stockholm University, SE-106 91 Stockholm, Sweden}

\author{A. Shekhter}
\email[email: ]{arkady.shekhter@gmail.com}
\affiliation{Los Alamos National Laboratory, Los Alamos NM 87545}

\begin{abstract} 
		The quasiparticle density of states in correlated and quantum-critical metals directly probes the effect of electronic correlations on the Fermi surface. Measurements of the nuclear spin-lattice relaxation rate provide one such experimental probe of quasiparticle mass through the electronic density of states.  
		By far the most common way of accessing the spin-lattice relaxation rate is via nuclear magnetic resonance and nuclear quadrupole resonance experiments, which require resonant excitation of nuclear spin transitions. 
		Here we report non-resonant access to spin-lattice relaxation dynamics in AC-calorimetric measurements. The nuclear spin-lattice relaxation rate is inferred in our measurements from its effect on the frequency dispersion of the thermal response of the calorimeter-sample assembly. We use fast, lithographically-defined nanocalorimeters to access the nuclear spin-lattice relaxation times in metallic indium from 0.3~K to 7~K and in magnetic fields up to 35~T. 
\end{abstract} 
\date{\today}
\maketitle
\setlength{\parindent}{0.5cm}
\setlength{\parskip}{0.1cm}

					\begin{figure}[t!!!!!!!!!!!]
								\begin{center}
							\includegraphics[width=0.85\columnwidth]{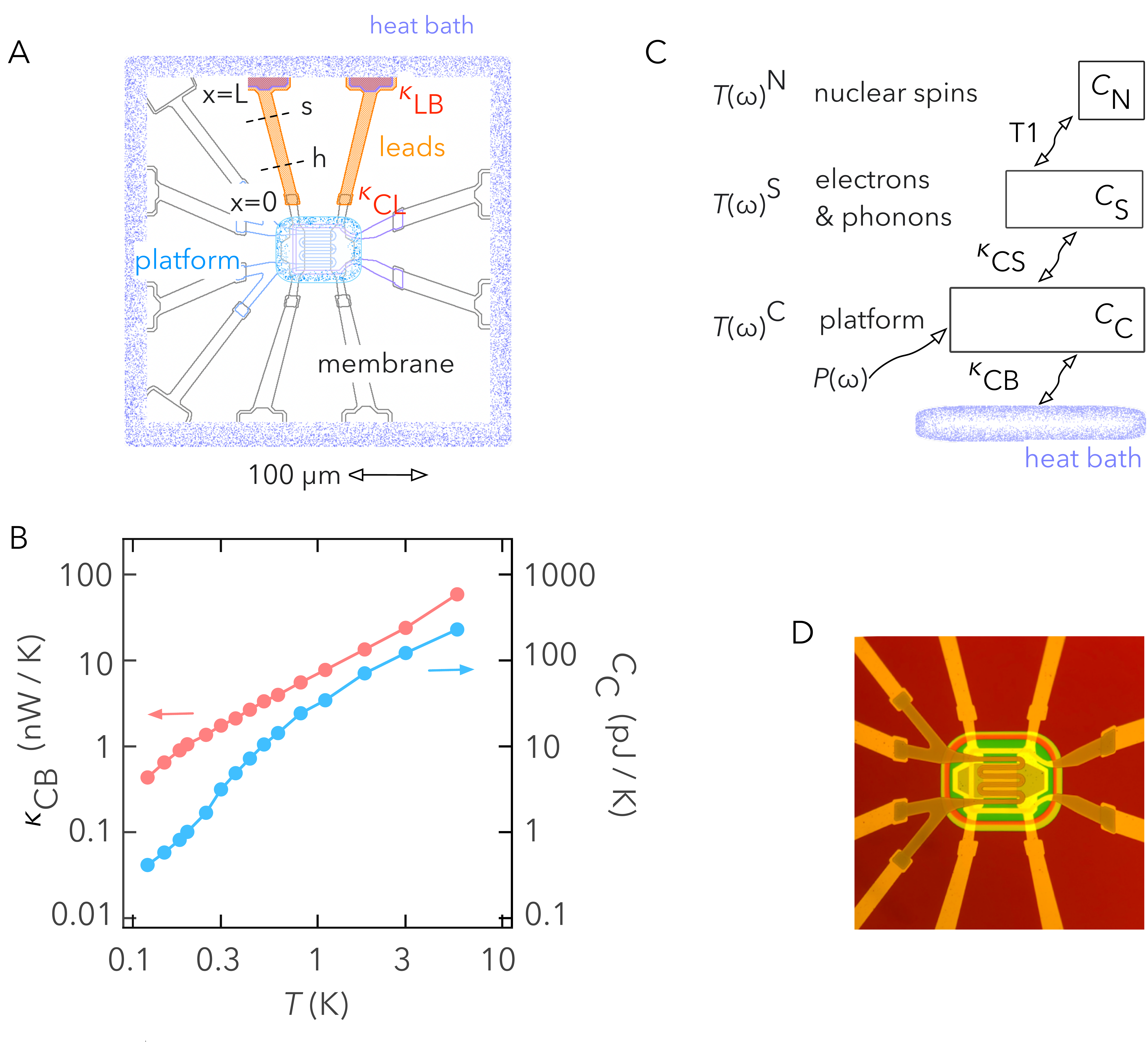}
								\end{center}
								\vspace{-0.8cm}
							\caption{  
									Thermal impedance spectroscopy with lithographic nanocalorimeters. 
								A. Sketch of the components of the nanocalorimeter. Colored elements indicate the heat bath (purple), chromium leads (orange), and calorimeter platform (blue) which contains the heater and the thermometer as well as the 1x1 mm SiN membrane (black). 
								The distance markers $x=0,L,s,h$ on the lead (highlighted in orange) illustrate the notation used in the discussion of the thermal susceptibility of a quasi-onedimensional object in Appendix~\ref{sec:xifunctions}. $κ_{\L\B}$ and $κ_{\C\L}$ indicate the thermal conductance of the leads-bath contact and lead-platform contact respectively used in the discussion in Appendix~\ref{sec:calorimeter-sample}. 
								B. Thermal conductance of calorimeter-heat-bath heat link $κ_{\C\B}$ and heat capacity of the calorimeter platform $C_{\C}$. The ratio of $C_{\C}/κ_{\C\B}$ is an indicator of the characteristic time of the calorimeter platform, ranging from 1 ms to 2 ms between 0.1~K and 1~K.  
								C. The heat flow diagram of the calorimeter-sample assembly. The sample is thermally coupled to the calorimeter platform via a thin layer of grease with contact conductance $κ_{\C\S}$. 
								D. Optical image of the nanocalorimeter. The gold-capped chromium leads are 400~$μ$m long, 35 $μ$m wide, and 60~nm thick. The 1 x 1 mm$^2$  SiN membrane is 150~nm thick \cite{Tagliati2012,Willa2017}. 
								}
								\label{fig:fig1} 
								\end{figure}

\section{Introduction}

In nuclear magnetic resonance (NMR) and nuclear quadrupole resonance (NQR) experiments, the nuclear spin transitions are excited resonantly by radio-frequency (RF) pulses. To measure the nuclear spin-lattice relaxation rate, the first RF pulse takes nuclear spins out of their equilibrium with the electrons on the Fermi surface and lattice excitations (phonons), and the second pulse probes their relaxation toward equilibrium \cite{Purcell1945, Purcell1951, Bloembergen1948, Abragam1961}. By applying the heat load on the lattice, nuclear spin transitions can be excited non-resonantly by the same spin-lattice interactions that are responsible for nuclear spin relaxation towards equilibrium. With fast calorimeters, one can resolve the dynamics of the heat flow between the lattice and the nuclear spins, as manifest in the delayed temperature response of the lattice to the heat load. 

Because of the large mismatch between nuclear- and electronic energy-level splittings in a magnetic field, the nuclear spin-lattice relaxation times in elemental metals are extremely slow compared to other microscopic timescales \cite{Abragam1961}, ranging from 5 ms in thallium at 1 K, 100 ms in indium and palladium, and 1 to 50 seconds for most other metals. These timescales can be accessed in small calorimeters. The slow spin-lattice relaxation dynamics determines the characteristic time delay between the temperature of the sample and the heat load on it through the thermal link to the calorimeter platform. Equivalently, the spin-lattice relaxation shows up as the characteristic time in the frequency dependence of the complex thermal impedance of the calorimeter-sample assembly. 

The specific heat of the nuclear spins (nuclear Schottky) and the specific heat of the lattice (electrons $+$ phonons) can be determined {\it independently} by such “thermal impedance spectroscopy” (TISP), because they are distinguished by their time-delayed response rather than by their magnitude. This presents a significant advantage, compared to the traditional, “static” specific heat measurements, where the two are superimposed. Compared to NMR and NQR measurements,  the non-resonant excitation of nuclear spins in TISP experiments puts less stringent requirements on the homogeneity of the internal and external magnetic fields. Here we report TISP measurement of the nuclear spin-lattice relaxation rate in metallic indium.

\section{Thermal impedance of the calorimeter-sample assembly}

Our lithographically-defined nanocalorimeter consists of a 150 nm-thick, 1x1 mm$^2$ SiN membrane with a 100x100 $μ$m$^2$ calorimeter platform at the center \cite{Tagliati2012, Willa2017}. The platform contains a calorimeter stage, heater, and thermometer, all in fast thermal contact with each other (Figure 1A,D). The thermal link to the heat bath is provided by gold-capped chromium leads (Figure 1A,D). The strength of the link is approximately 10~nW/K at 1 K and increases approximately linearly with the temperature between 0.1~K and 10 K (Figure 1A). The heat capacity of the calorimeter platform is 10 pJ/K at 1~K and increases approximately linearly with temperatures between 1~K and 10~K (Figure 1B). The nominal characteristic time of such a calorimeter, given by the ratio of the heat capacity of the calorimeter platform and the thermal conductance of the thermal link, is 1~ms. On a millisecond time scale, the calorimeter stage, thermometer, and heater act as a monolithic calorimeter platform with uniform temperature $T(t)^{\C}$ measured by the thermometer (Figure 1A,D). A metallic indium sample with a mass of 1.7 $μ$g (15 nmol) is mounted on the calorimeter platform using a thin layer of grease. 

				\begin{figure}[t!!!!!!]
					%\captionsetup{format=plain, justification=default, width=.8\textwidth}
					\begin{center}
					\includegraphics[width=0.7\columnwidth]{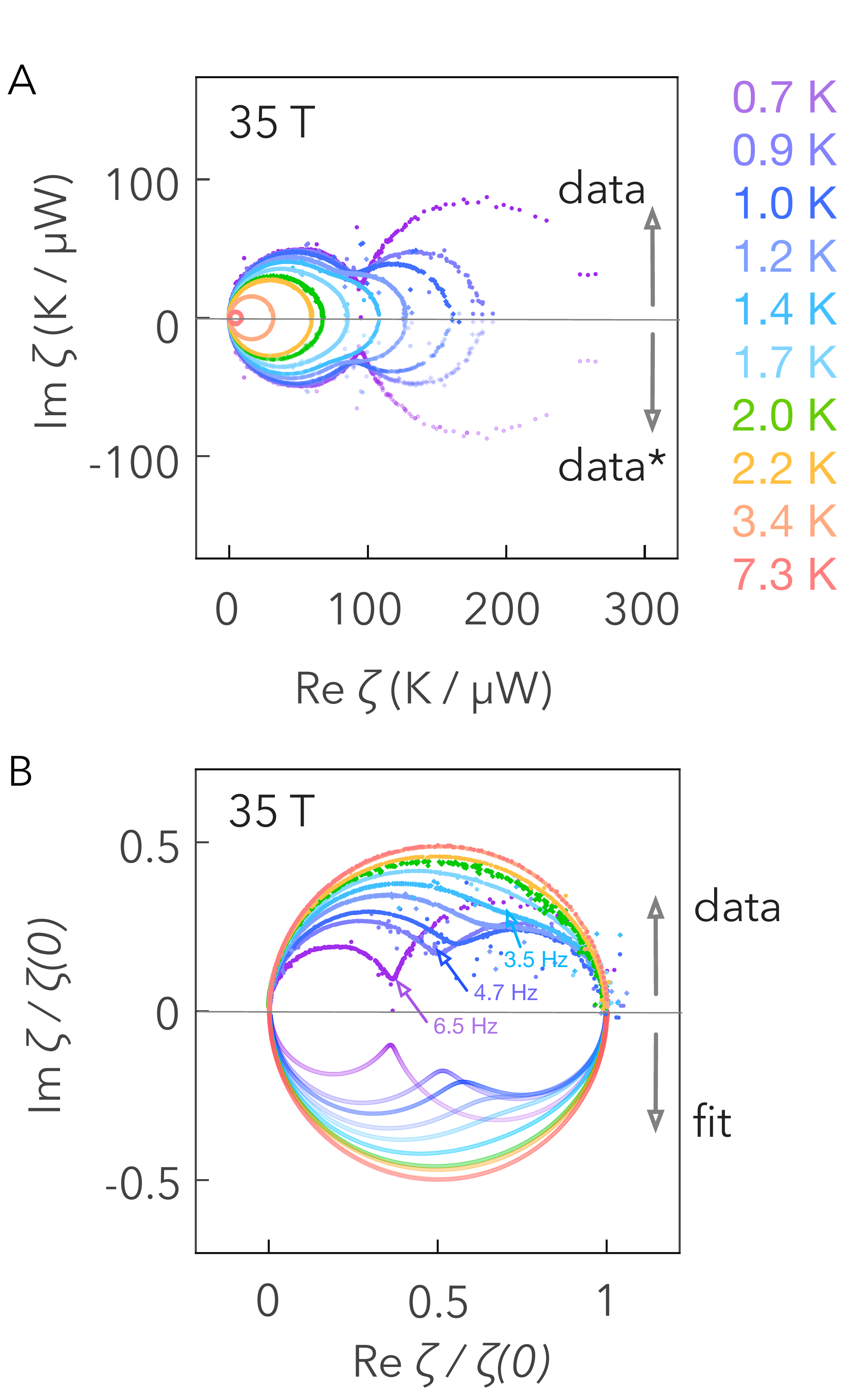}
					\end{center}
						\caption{ %\scriptsize 
					The thermal impedance of the calorimeter-sample assembly in the complex plane of $ζ$.
					A. Measured thermal impedance $ζ(ω)$ at 35~T  at temperatures in the range from 0.7~K  to 8~K. The upper half $\Imζ> 0$, shows the observed thermal impedance. The lower half, $\Imζ < 0$, is added as a guide for the eye to represent the thermal impedance at negative frequencies, $ζ(-ω) = ζ^*(ω)$.
					B. The normalized thermal impedance $ζ(ω) / ζ(ω=0) $ where $ζ(ω=0) = 1/κ_{\C\B}$. The upper half of the plot shows the observed normalized thermal impedance. The lower half shows the fit to Eq.~(\ref{eq:theR}) with fitting parameters discussed in Figure 3A. 
						} 
					\label{fig:fig2ab} 
					\end{figure} 

We drive an oscillating heat load, $P(ω)$ (Figure 1C), and measure the complex (both in-phase and out-of-phase) temperature response of the calorimeter platform, $T(ω)^{\C}$. This defines a complex thermal impedance \cite{Richardson2018, Pobell2007} of the calorimeter-sample assembly, $ζ(ω)^{\C} = T(ω)^{\C} / P(ω)$. Figure 2A shows the observed thermal impedance of the metallic indium sample in the frequency range of 10 mHz to 1 kHz plotted in the complex plane of $ζ^{\C}$. Figure 1C shows the different components that make up the thermal response of the calorimeter. The thermal impedance corresponding to the heat flow diagram in Figure 1C is given by 
\begin{align}\label{eq:theR} 
\frac1{	ζ(ω)^{\C}} =  κ_{\C\B}   -i ω C_{\C} 
 + \frac{ -i ω \! \para{ C_{\S}  + \frac{  C_{\N} }{ -i ω T\!1 + 1 } } κ_{\S\C}  }{  -i ω\!\para{ C_{\S}  +  \frac{ C_{\N} }{ -i ω  T\!1+ 1 } } + κ_{\S\C}     } \,,
\end{align}
where $κ_{\S\C} $ is the thermal conductance of the contact between the calorimeter platform and the sample, $κ_{\C\B}$ is the thermal conductance from the calorimeter platform to the heat bath, $T1$ is the nuclear spin-lattice relaxation time, and $C_{\C, \S, \N}$ are heat capacities of the calorimeter platform, and electron+phonon and nuclear spin subsystems in the sample, respectively. Eq.~(\ref{eq:theR}) faithfully describes the heat circuit in Figure 1C below 3~kHz, set by the thermal diffusion time across the leads, 50~$μ$s at 1~K.  At higher frequencies, the frequency dispersion of the thermal link  $κ_{\C\B}(ω)$ needs to be accounted for (Appendix \ref{sec:calorimeter-sample}).

The thermal impedance in Eq.~(\ref{eq:theR}) is a superposition of three complex Lorentzians, $ζ(ω) = ∑_{i=1,2,3} \, {A_i}/{[-iω + 1/m_i]}$, all centered at zero frequency. The amplitudes $A_{1,2,3}$ are real and characteristic times, denoted $m_{1,2,3}$, are all positive, by causality. The temperature dependencies of $m_{1,2,3}$ obtained from the data in Figure 2 are shown as gray lines in Figure 3A. The two longer characteristic times, $m_{1}$ and $m_2$, span the interval from 1~sec  to 1~ms in the temperature interval from 0.2~K to 7~K. The measurement frequency range encompasses the two longer characteristic times  $m_{1}$ and $m_2$ in this temperature interval. 

We can define three timescales using parameters in Eq.~(\ref{eq:theR}), $τ_{\S} = C_{\S}/κ_{\C\B} $, $τ_{\C}=C_{\C}/κ_{\C\B} $, and $ τ_{\N} = C_{\N} / κ_{\C\B}$. Figure 3A shows the temperature dependence of $τ_{\S}, τ_{\C}$, and $τ_{\N}$ determined by the fit of the observed frequency dependence to Eq.~(\ref{eq:theR}) at each temperature and magnetic field (Appendix \ref{sec:fitting}). 
%	The longest characteristic time $m_1$ of the calorimeter-sample assembly is approximately set by the maximum of $τ_{\N}, τ_{\S},$ and $T1$, i.e. $m_1 \approx \max\{ τ_{\N}, τ_{\S}, T1\}$. Between 0.7~K and 3~K, $m_1$ is approximately equal to nuclear spin-lattice relaxation time T1. 

At higher temperatures, the sample and the calorimeter platform are in thermal equilibrium. In this low-frequency regime, the thermal impedance in Eq.~(\ref{eq:theR}) reduces to a single-characteristic-time form used in AC-calorimetric measurements \cite{Sullivan1968, Tsui1988, Shayegan1991, Tagliati2012, Willa2017}, ${ζ(ω)^{\C}} = \fra1{[κ_{\C\B} -  i ω (  C_{\C}\!+C_{\S} \!+\!C_{\N}) ]}$. At 7~K, this low-frequency regime extends up to about 20 Hz, which is the inverse of the second-largest characteristic time at that temperature (Figure 3A). The plateauing of the amplitude at 7 K below about 1 Hz (red curve in Figure 3B) corresponds to the longest characteristic time $m_1$ at 7 K (Figure 3A). This low-frequency regime corresponds to a large circle in the complex plane of $ζ^{\C}$, defined by the Lorentzian ${A_1}/{[-iω + 1/m_1]}$ with the longest characteristic time $m_1$ shown in Figure 2B where we plot the normalized value of the thermal impedance $ζ(ω)^{\C}/ζ(ω=0)^{\C}$.

\begin{figure}[t!!!!!!]
		%\captionsetup{format=plain, justification=default, width=.8\textwidth}
		\begin{center}
		\includegraphics[width=0.99\columnwidth]{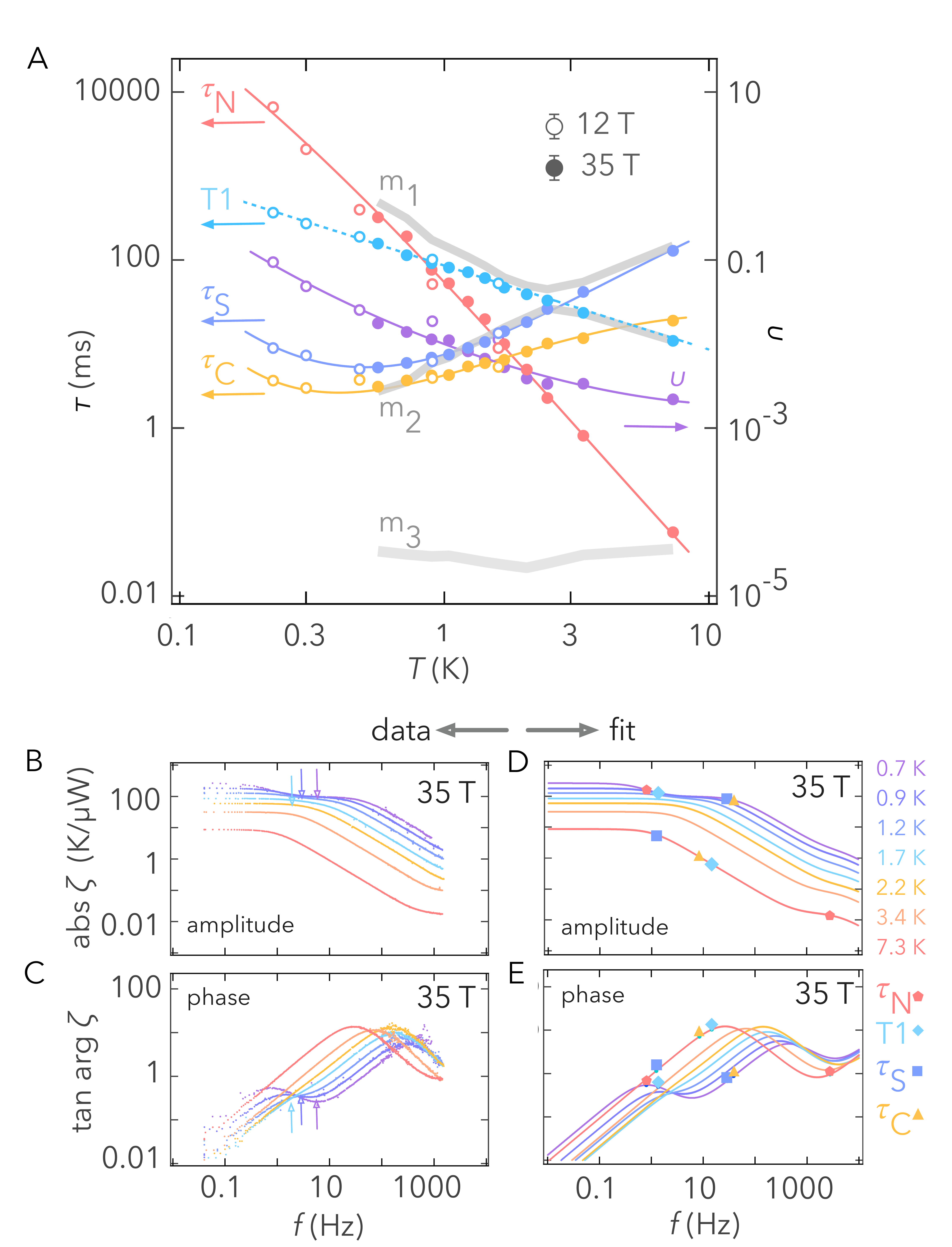}
		\end{center}
		\caption{
		Fitting the thermal impedance of the calorimeter-sample assembly.
	A. The temperature dependence of the fitting parameters $τ_{\S} = C_{\S}/κ_{\C\B} $, $τ_{\C}=C_{\C}/κ_{\C\B} $, and $ τ_{\N} = C_{\N} / κ_{\C\B} $, the nuclear spin-lattice time, $T1$, and a dimensionless ratio $\ga =  κ_{\C\B} / κ_{\S\C} $, determined by the fit to Eq.~(\ref{eq:theR}). The solid circles represent measurements in the resistive magnet at 35~T on a 15 nmol-size sample. The open circles represent measurements in the superconducting magnet at 12~T on a 7.8 nmol sample. The nuclear Schottky for the 12~T measurements is scaled by a factor (35/12)$^2$. Thick gray lines represent the characteristic times $m_1$, $m_2$, and $m_3$ of the calorimeter-sample assembly determined by the observed thermal impedance via $ζ(ω) = ∑_{i=1,2,3} \, {A_i}/{[-iω + 1/m_i]}$. The dashed cyan line indicates the nuclear spin-lattice time T1 measured in NQR experiments \cite{Maclaughlin1966, Maclaughlin1971}. Solid lines tracing $τ_{\S,\N,\C}$ and   $\ga$ are guides for the eye. 
		B,C. Frequency dependence of the polar components (amplitude and phase) of the observed thermal impedance in the frequency range of 10mHz to 1 kHz.  
		D,E. Frequency dependence of the polar components of $ζ(ω)$ in Eq.~(\ref{eq:theR}) with the best-fit parameters from A. The values of parameters $τ_{\C,\S,\N}$ and T1 are indicated by markers at $f = 1/(2\pi τ_{\C,\S,\N})$ on top of 7.3 K and at 0.7 K frequency scans. 
		} 
		\label{fig:fig2cdefg} 
		\end{figure}

As we lower the temperature, the internal thermal equilibrium between the calorimeter platform and the sample, and between the electron-phonon and nuclear-spin subsystems of the sample, breaks down. The multiple-relaxation-time character of the frequency dependence of the thermal impedance in this regime is evident in the ``multicircle'' geometry of the frequency traces of $ζ(ω)$ in the complex plane (Figures 2A,B). At 0.7 K the two resonances in $ζ(ω)^{\C}$ are clearly visible in the amplitude plot (Figure 3B). The plateau below 30 Hz corresponds to a broad resonance with a characteristic time of 5 ms, the intermediate characteristic time $m_2$ in Figure 3A at 0.7~K, which is close to $τ_{\C}$ and $τ_{\S}$ at that temperature. The plateau in the amplitude below 0.3 Hz (purple curve in Figure 3B) corresponds to the longest characteristic time of 0.5 sec in Figure 3A at 0.7K. 

\section{ TISP measurements of indium metal} 
	
Figure 4A shows the specific heat of the nuclear spin (in red) and lattice (electrons $+$ phonons,   in blue) subsystems, as extracted from the data shown in Figure 2 using the model given by Eq.~(\ref{eq:theR}). The specific heat of metallic indium, measured in Ref.~\onlinecite{oneal1964}, is shown as a dashed blue line in Figure 4A. Naturally-occurring indium has two isotopes: $^{115}$In (95.7\%) and $^{113}$In  (4.3\%). Both isotopes have nuclear spin $J=9/2$ and $g$-factors differing by 0.2\%, $^{115}g_{\N}\!=\!+\!1.231$ and $^{113}g_{\N}\!=\!+\!1.229$ \cite{Stone2014}. At low magnetic fields, $μ_{\N} B ≪ k_{\B} T$, the nuclear Schottky per mol is  $c_{\N} = \frap13J(J\!+\!1) N_{\A}k_{\B} \para{\fra{g_{\N} μ_{\N} B }{ k_{\B} T }}^2$ where $μ_{\text{N}}=32.5$ neV/T is the nuclear magneton. For elemental indium, this is shown as a dashed red line in Figure 4A. 

The nuclear spin-lattice interactions in metallic indium were studied previously in  NQR \cite{Maclaughlin1966, Maclaughlin1968, Maclaughlin1971, Maclaughlin1973} and NMR \cite{Adams1966,Hammond1966} experiments. Figure 4B shows the nuclear spin-lattice relaxation rate $1/T1T$ determined by the fits of Eq.~(\ref{eq:theR}) to the data in Figure 2. The dashed line shows the value of $T1T = 0.086$~s~K from NQR experiments \cite{Maclaughlin1966, Maclaughlin1971}. 

\section{The thermodynamics of the nuclear spin relaxation.}

We now briefly discuss the thermodynamic description of nuclear spins\cite{Casimir1938,VanVleck1940,   Hebel1963,Abragam1961}, as it relates to TISP experiments. In a metal, the spin temperature is established by spin-spin interactions on the timescale of the spin-spin relaxation time, T2 of the order of  1~-~100 $μ$s \cite{Bloembergen1953, Abragam1961}, faster than T1 in a typical metal. In indium metal, T2 $=$ 100 $μ$s \cite{Maclaughlin1971}. Thus, the nuclear spin temperature is well-defined on the timescale of 1 ms. 
The non-uniform spatial dependence of the nuclear spin- and lattice temperatures do not need to be similar to each other, despite the fact that the spin-lattice interaction is local. This is because the heat can flow ``laterally" between the nuclear spins mediated by spin-spin interaction and across the lattice, mediated by heat diffusion. 
The (lattice) heat diffusion time across the cuboid-shaped sample of mass of 1.7 $μ$g is short, about 10~ns (Appendix \ref{sec:calorimeter-sample}). Therefore, both the lattice $T_{\S}$ and  spin $T_{\N}$ temperatures are uniform across the sample in our measurements, as assumed in Eq.~(\ref{eq:theR}). 

Different isotopes of the same nucleus will be out of equilibrium with each other when an oscillating heat load is applied to the lattice because spin-spin relaxation is ineffective in establishing the common temperature for isotope species when energy-matching conditions between them are broken. 		
Indium, however, is peculiar in this respect: the $g$-factors of its two naturally occurring isotopes, $^{113}$In and $^{115}$In, differ by only 0.2\%. 
At low magnetic fields, the nuclear-spin energy levels of the two isotopes effectively overlap due to finite-width effects. The nuclear-spin energy states are broadened by inelastic energy exchange mediated by either the spin-spin or the spin-lattice interactions. In metals, the energy transfer mediated by the spin-spin interactions dominates these effects\cite{Abragam1961, Hebel1963, Bloembergen1959}. In metallic indium, the width of the nuclear energy states, $∼ ℏ/T2$,  is comparable to the energy mismatch of its two isotopes in applied magnetic fields of 10~T or less.

		\begin{figure}[t!] 
					\begin{center}
					\includegraphics[width=0.99\columnwidth]{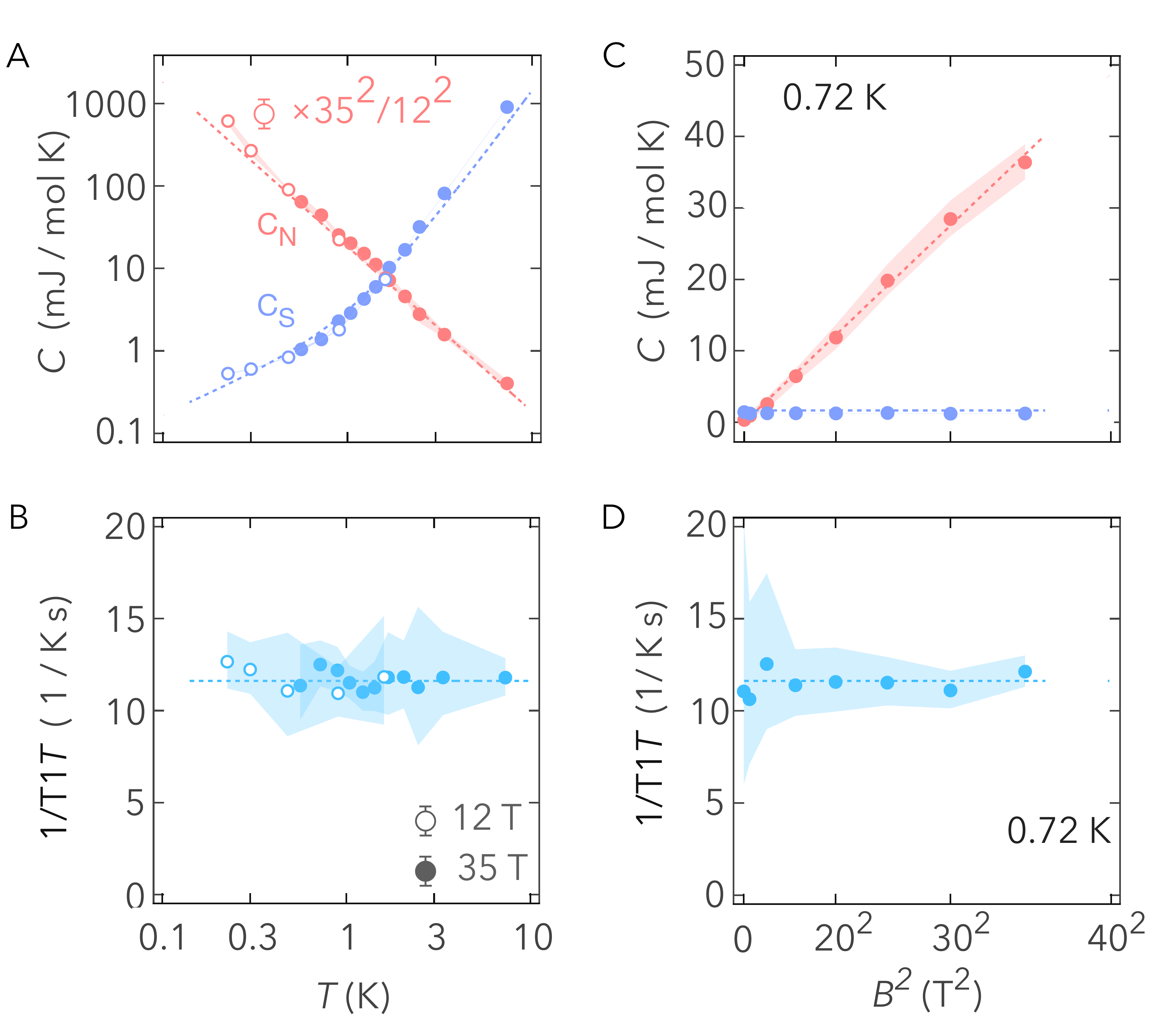}
					\end{center}
					\caption{  
					Temperature and magnetic field dependence of the spin-lattice relaxation rate and nuclear- and lattice heat capacity. 
					A. Magnetic field dependence of the nuclear Schottky $C_{\N}$ (in red) and electronic specific heat $C_{\S}$  (in purple) at 0.7~K.  The magnetic field axis scales as field-squared. The shaded region around the line represents the error bars (Appendix \ref{sec:fitting}).
					B. The nuclear spin-lattice relaxation rate 1/T1T vs field.  
					C. Temperature dependence of the nuclear Schottky $C_{\N}$ (per mol) and the electronic ($+$ phonon) specific heat $C_{\S}$ (per mol) at 35~T  (solid circles) and 12~T  (open circles). For the sake of comparison, the nuclear Schottky at 12~T is scaled up by a factor of $(35/12)^2$. The dashed blue line indicates heat capacity measured in Ref.~\onlinecite{oneal1964},  $c_S = (1.69 \text{ mJ / mol K}^2) \, T + (1.43 \text{ mJ / mol K }^4) \,T^3$. The dashed red line indicates the expected magnitude of the nuclear Schottky in indium metal, $c_N =   (0.015 \text{ mJ K / mol T}^{2}) \; B^2 / T^2$.	
					D. Temperature dependence of the nuclear spin-lattice relaxation rate 1/T1T at 35~T (solid  circles) and 12~T  (open circles). 		
						}
					\label{fig:fig3}
					\end{figure}

Regardless, the detection of partial thermal equilibrium between indium isotopes in metallic indium requires a sub-percent-level of determination of T1, beyond our current sensitivity (Figure 4B). The partial equilibrium between different isotopes (or different lattice sites with different Knight shifts) might be an important consideration in other systems (Appendix \ref{sec:crossrelaxation}). 

Finally, the spin-lattice time T1 is defined in Eq.~(\ref{eq:theR}) via effective ``contact’’ thermal conductance between the nuclear-spin and the lattice, $κ_{\S\N}^{\eff} = C_{\N} / T1$. To establish a connection between the TISP measurements and the NMR and NQR experiments \cite{Abragam1961, Hebel1963, Bloembergen1959}, we consider the energy exchange between the nuclear spins and the lattice. 
The approach to thermal equilibrium is governed by the relaxation dynamics of the entropy \cite{Onsager1930, Onsager1931},  $\fra{d(S_{\N}+S_{\S})}{dt} = q^{\NfromS} \para{ \fra1{T_{\N}}-\fra1{T_{\S}}}$, where heat flux from the electron-phonon subsystem to the nuclear spins, $q^{\NfromS}= T_{\N}\frap{dS_{\N}}{dt} =  -T_{\S}\frap{dS_{\S}}{dt}$, is equal to the rate of change of energy of nuclear-spin subsystem, $q^{\NfromS} = \fra{dQ_{\N}}{dt}$. Following Refs.~(\onlinecite{Onsager1930, Onsager1931}), we chose $x = dQ_{\N} = C_{\N} dT_{\N}$ as a measure of the deviation from complete equilibrium and $X = \fra1{T_{\N}} - \fra1{T_{\S}}$ as the corresponding thermodynamic force. Then the rate of change of $x$ is proportional to the thermodynamic force, $\fra{dx}{dt} =  Γ \, X$. Rewriting $\fra{dx}{dt}= C_{\N}\; \fra{dT_{\N}}{dt} $ as $ - T_{\N}^2C_{\N} \,\times\,  \fra{\!d\frap{1}{T_{\N}}\!}{dt} $ we obtain the rate equation for the nuclear temperature 
\begin{align} 
	\frac{d\frap{1}{T_{\N}}}{dt} = - \frac1{T1}\para{ \frac1{T_{\N}} - \frac1{T_{\S}}}\,,
\end{align} 
where $1/T1 = Γ/(T_{\N}^2C_{\N})$ is the nuclear temperature relaxation rate. When temperature oscillations are weak, $\delta T \ll T$, this equation sets the effective contact thermal conductance between the nuclear spins and the lattice to $C_N/T1$. This establishes the equivalence of our definition of T1 in Eq.~(\ref{eq:theR}) and that in magnetic resonance experiments \cite{Abragam1961,Hebel1963}. 

The nuclear spin-lattice relaxation indicates the spin-flip dynamics of electrons in the host material \cite{Purcell1945, Purcell1951, Bloembergen1948}. In conventional metals and in liquid  $^3$He below 1~K  \cite{Leggett1975}, all spin flips occur on the Fermi surface \cite{Bloembergen1953, Abragam1961, Korringa1950} and the nuclear spin-lattice relaxation rate indicates the quasiparticle density of states \cite{Korringa1950} as well as the static Fermi liquid renormalization factors \cite{Landau1956, Leggett1968, Leggett1975, AndersonRedfield1959, Morya1963}. Beyond conventional metals, nuclear spin-lattice relaxation probes the correlation dynamics and its impact on the Fermi surface, as well as Fermi liquid renormalization factors not included in mass renormalization. One advantage of TISP measurements in this broader scientific context is that the nuclear spin-lattice relaxation rates can be measured simultaneously with electronic specific heat, thus providing two independent and complementary ways to access the quasiparticle mass or the density of states on the Fermi surface in the same measurement. 

{\it Acknowledgments.} 
We thank Stuart Brown, Michihiro Hirata, and Aharon Kapitulnik for illuminating discussions. 
Resistive 35 T magnet measurements were performed at the National High Magnetic Field Laboratory, which is supported by the National Science Foundation Cooperative Agreement No. DMR-1644779 and the State of Florida. 
Work at Los Alamos National Laboratory is supported by the NSF through DMR-1644779 and the U.S. Department of Energy. 
A.S. acknowledges support from the DOE/BES ``Science of 100 T'' grant. 
A.S. acknowledges the hospitality of the Aspen Center for Physics, where a portion of this work was performed. Aspen Center for Physics is supported by National Science Foundation grant No. PHY-1607611. 
B.J.R. acknowledges funding from the National Science Foundation under grant no. DMR-1752784. 
A.K. and A.R. acknowledge support by the Swedish Research Foundation, D. Nr. 2021-04360.

\appendix

\section{Thermal impedance.}
\label{sec:methods}

In lithographically defined nanocalorimeters \cite{Tagliati2012, Willa2017},  the calorimeter stage, heater, and thermometer are in tight thermal contact  (Figure~1). The complete thermal signature of the calorimeter-sample assembly (Figure 1) is obtained by measuring its thermal impedance $ζ(t-t')^{\C}$ \cite{Richardson2018, Pobell2007}, which describes the temperature change $δT(t)^{\C}$ of the thermometer at time $t$ in response to the heat influx at the heater $δ P(t')$ at an earlier time~$t'$,  
\begin{align}\label{eq:thermalimpedance}
	 δT(t)^{\C} \! =& \!\!∫\limits_{\!\!\!\!\!\!\!-∞}^{\,\,t}  \! dt' \, ζ(t\!-\!t')^{\C}  P(t') \,, \;\; 
	 T(ω)^{\C}   \!\! = \! ζ(ω)^{\C}  P(ω)  \,.
\end{align} 
The second equation defines the thermal impedance in the frequency domain measured directly in AC calorimetry. The time-dispersion of $ζ(t-t')^{\C}$ encapsulates an internal memory of the history of heat injection over a finite time interval into the past. To describe components of the thermal circuit with finite thermal memory we also use dimensionless thermal susceptibilities, $Ξ(ω)\!=\!-iω C ζ(ω) $, which describe the frequency-dispersed response to the heat influx of the {\it rate} of temperature change, $-iωC T(ω) =  Ξ(ω)q(ω)$. For thermally isolated components it approaches unity in the low-frequency limit, $Ξ(ω→0)=1$.

On millisecond time scales, the calorimeter stage, thermometer, and heater act as a monolithic ``calorimeter platform'' with temperature $T(t)^{\C}$ measured by the thermometer. 

At very low frequencies, the calorimeter platform ($C_{\C}$) is in thermal equilibrium with the sample (electron-phonon subsystems ($C_{\S}$) and nuclear spin ($C_{\N}$)), 
$-iω(C_{\S} \!+\! C_{\C}\!+\!C_{\N}) T(ω)^{\C}   =  P(ω)  +	 q(ω)^{\CfromB} $, 
where 
$q(ω)^{\CfromB} =  - κ_{\C\B} T(ω)^{\C}$ 
is the heat flux from the thermal bath into the calorimeter platform and $κ_{\C\B}$ is its heat conductance (Figure 1). In this limit the thermal impedance is characterized by a single characteristic time, 
\begin{align}\label{eq:singlepole} 
	ζ(ω) = \frac1{ -iω( C_{\S} \!+\! C_{\C}\!+\!C_{\N}) + κ_{\C\B} } \,.
\end{align} 
determined by the ratio of the total heat capacity $C_{\S} \!+\! C_{\C}\!+\!C_{\N}$ and $κ_{\C\B}$ (Figure 1D). Thus, the total heat capacity $C_{\S} \!+\! C_{\C}\!+\!C_{\N}$ and $κ_{\C\B}$ can be determined from the measurement of $ζ(ω)$ at a single frequency, typically chosen  near $\text{arg}ζ(ω)=45^{∘}$ \cite{Sullivan1968, Tsui1988, Shayegan1991, Tagliati2012, Willa2017}.  At higher frequencies, the sample and the calorimeter platform are out of equilibrium, and single-characteristic time no longer faithfully represents the thermal response of the calorimeter.   

%\subsection{thermal impedance of the heat circuit in Figure 2A.}

To access nuclear relaxation rates, one needs to analyze the thermal impedance of a calorimeter with its components out of equilibrium. 

\section{Thermal impedance of the calorimeter-sample assembly.}
\label{sec:calorimeter-sample}

The heat balance of the calorimeter platform (Figure 1AC) is described by 
\begin{align}\label{eq:platform} 
-i ω C_{\C} T(ω)^{\C}     %  \notag\\
\!\! =\!  P(ω)  % &
       \!- \! q(ω)^{\SfromC}  	
\!\! - \! q(ω)^{\LfromC}  
\!\! - \! q(ω)^{\MfromC} \!\! \,.
\end{align}
The right-hand side lists all the heat sources of the calorimeter platform, including the heat flux from the membrane $-q(ω)^{\MfromC} $, the electric leads $-q(ω)^{\LfromC}$, and the sample, $-q(ω)^{\SfromC}$, determined self-consistently by the temperature $T(ω)^{\C}$ of the calorimeter platform. 

%\subsection{Heat link to the heat bath.}

The heat fluxes across the leads and across the membrane $-q(ω)^{\LfromC} $ and $-q(ω)^{\MfromC} $ act together to provide the heat link to the heat bath (Figure~1A,C)\,, 
\begin{align}\label{eq:leadmem} 
q(ω)^{\LfromC} + q(ω)^{\MfromC} = κ_{\C\B}(ω) \; T(ω)^{\C} \,   ,
\end{align} 
where $κ_{\C\B}(ω) $ is the frequency-dependent heat conductance from the calorimeter platform to the heat bath. 

The heat conductance through leads can be understood in terms of the series-connected heat conductances of the lead-platform contact, $κ_{\L\C} $, the lead-heat-bath contact, $κ_{\L\B}$, and the leads themselves, $k_{\L} A/L$ (where $A$ is the crosssectional area of the leads and $L$ is their length), $1/κ_{\C\B} = 1/κ_{\L\C} + 1/κ_{\L\B} + 1/(k_{\L} A/L)$ (Figure 1A). Of the three, the heat conductance of the leads $k_{\L} A/L$ is the weak link, because both $κ_{\L\C}$ and $κ_{\L\B}$ are controlled by the thermal flow across wide metallic surfaces in the lithographic assembly. The combined heat conductance of the gold-capped chromium leads ($κ_{\L}=$ 5~mW/cm~K Figure 1AC) is 10~nW/K at 1~K, comparable to the measured value of $κ_{\C\B}$ at this temperature (Figure 1B). The SiN membrane accounts for about 0.3~nW/K ($κ_{\M}=$ 1~mW/cm~K, Figure 1A). The heat diffusion time across the 400 $μ$m-long gold-capped chromium leads, $τ_{\L} =  (c_{\L} / κ_{\L} ) L^2 $, is estimated as $L^2/(v_F ℓ/3) = $ 50 $μ$s at 1~K, where $ℓ$ is the mean free path in chromium at 1~K \cite{LandauVol10, Tanuma2011}.  Therefore, $κ_{\C\B}$ is independent of frequency below about 3~kHz. 

The same conclusion is reached in a detailed analysis of the heat balance in the leads (Figure 1AC in the text), 
\begin{align}\label{eq:heateqKsKm} 
-i ω C_{\L} 
		\matr{T(ω)_{x=0} \\  T(ω)_{x=L}  }^{\L} 
	 	\!\!\!= \!\! & \, 
	 	\matr{ 
	 	ξ(00;ω)  &  ξ(0L;ω) \\ 
	 	ξ(L0;ω)  &  ξ(LL;ω)  
	 	}^{\L} 
	 	\matr{ q(ω)_{x=0}  \\ q(ω)_{x=L}  }^{\L} \,, \notag\\
 q(ω)_{x=0} ^{\L}  
 \; =& \; -κ_{\L\C}\;[\; T(ω)^{\L}_{x=0} - T(ω)^{\C}  \;] \,,\notag\\
 q(ω)_{x=L}^{\L} =& - κ_{\L\B} \; T(ω)^{\L}_{x=L} \,.
\end{align} 
The first equation determines the temperature $T(ω)^{\L}_x$ at the $x=0,L$ ends of the leads when two heat sources, $q(ω)_{x=0}$ and $q(ω)_{x=L} $, act simultaneously. $ξ(sh;ω)$ is the dimensionless thermal susceptibility of the bar-like object (Appendix \ref{sec:xifunctions}) \cite{Chandrasekhar1960},  describing temperature at point $s$ when heat is applied at point $h$ in the bar (Figure 1A), 
$ -i ω C_{\L} T(ω)^{\L}_{x=s}  = \; ξ(sh,ω) \; q(ω)^{L}_{x=h}$.

The second line in Eq.~(\ref{eq:heateqKsKm}) determines the heat flux $q(ω)^{\L}_{x=0}$ at the calorimeter end of leads in terms of the temperature difference at that contact point ($κ_{\L\C}$ is the contact conductance). The third equation determines the heat flux $q(ω)^{\L}_{x=L}$ at the heat bath end of leads in a similar way ($κ_{\L\B}$ is the contact conductance). 

Solving Eq.~(\ref{eq:heateqKsKm}) for $q(ω)^{\LfromC} = q(ω)_{x=0}^{\L}$ we find 
\begin{align}\label{eq:heatlink}
q(ω)^{\LfromC} & = \; T(ω)^{\C}  \frac{-iω C_{\L}}{ -iω\;(C_{\L}/ κ_{\C\L} ) + Ξ(ω)^{\L} } \,, \notag\\ 	 
 \text{where} & \quad   Ξ(ω)^{\L}  =\;  
 \frac{ - iω τ_{\L}  \!\!-\! i ω \; (C_{\L}/κ_{\L\B})\;   ξ(00;ω)^{\L}   }{  -i ω \; (C_{\L}/κ_{\L\B})   +   ξ(00;ω)^{\L} }  \,.
\end{align}
The  factor multiplying $T(ω)^{\C}  $ on the first line is equal to $κ_{\C\B}(ω)$. In the low-frequency limit, $ωτ_{\L} ≲ 1$, the thermal susceptibility of the leads approaches unity (Appendix \ref{sec:xifunctions}),  $ξ(00;ω) ^{\L} =1$, and thermal susceptibility in the second line approaches $Ξ(ω≪\!1/τ_{\L})^{\L}  \!=\!  -iω [τ_{\L} + (C_{\L}/κ_{\L\B})]$. Therefore, when $ωτ_{\L} ≲ 1$, all frequency factors in the  heat conductance $κ_{\C\B}(ω)$ cancel out, that is,  $κ_{\C\B}$ is frequency independent at frequencies below $1/τ_{\L}$.

%\subsection{Heat balance on the sample.}

The sample-platform heat flux $-q(ω)^{\SfromC} $ is determined by the heat balance for the sample (Figure 1C) 
\begin{align} \label{eq:samplebalance}
	 - i ω C_{\S} T(ω)^{\S} = &\, \,  Ξ(ω)^{\S} \, q(ω)^{\SfromC} - q(ω)^{\NfromS} \,, \notag\\
	 q(ω&)^{\SfromC}  =  - κ_{\S\C} [ \, T(ω)^{\S} - T(ω)^{\C} \, ] \,,
\end{align} 
where $κ_{\S\C}$ is the heat conductance of the contact between the sample and calorimeter platform. $Ξ(ω)^{\S}$ is the thermal susceptibility for the sample at the calorimeter contact end. The sample has a standing-bar geometry, therefore, $Ξ(ω)^{\S} = ξ(00;ω)$ (Appendix \ref{sec:xifunctions}).  At 1~K  the heat diffusion time $τ_{\S} = (c_{\S} / κ_{\S} ) L^2$ across the height ($L =$ 30 $µ$m) of the sample is 10~ns. Therefore, in this measurement the thermal impedance  $Ξ(ω)^{\S}$ is frequency-independent, $Ξ(ω ≪ 1/τ_{\S})^{\S} =1$. 

%\subsection{Heat balance of the nuclear spins.}

$q(ω)^{\NfromS}$ in Eq.~(\ref{eq:samplebalance}) is the heat flux from nuclear spins to the sample. The heat balance of the nuclear spins is described by 
\begin{align}\label{eq:spinlattice} 
	 - i ω C_{\N} T(ω)^{\N} & =  q(ω)^{\NfromS}  \,, \notag\\
		q(ω&)^{\NfromS}  =  - \frac{C_{\N}}{T1} [ T(ω)^{\N} - T(ω)^{\S} ] \,, 
\end{align}
where $T(ω)^{\N} $ is the temperature of the nuclear spin subsystem, and $T1$ is the spin-lattice relaxation time \cite{Abragam1961, Hebel1963,Lenk1979}.  Solving Eq.~(\ref{eq:spinlattice}) for $q(ω)^{\NfromS}$ in terms of electron-phonon temperature $T(ω)^{\S}$, Eq.~(\ref{eq:samplebalance}) for $q(ω)^{\SfromC}$ in terms of $T(ω)^{\C}$, and combining with Eqs.~(\ref{eq:platform},\ref{eq:leadmem}) we finally obtain the thermal impedance of the calorimeter, $ζ(ω)^{\C} = T(ω)^{\C}/P(ω)$, 
\begin{align}\label{eq:theRm}
\frac1{	ζ(ω)^{\C}}  
\!=\!  -i ω C_{\C} \!+\! κ_{\C\B} \!
\!+\!\frac{ -i ω \! \para{ \!C_{\S} \! +\! \frac{  C_{\N} }{ -i ω T\!1 + 1 } \!} κ_{\S\C}      }{ -i ω\! \para{\! C_{\S} \! +\!\!  \frac{ C_{\N} }{ -i ω T\!1+ 1 }\! } \!+\! κ_{\S\C}     } \,.
\end{align}

The frequency sweeps are fitted to the model, Eq.~(\ref{eq:theRm}), using six parameters, $κ_{\C\B} $,  $τ_{\S} = C_{\S}/κ_{\C\B} $, $τ_{\C}=C_{\C}/κ_{\C\B} $ , $T_1$, $ τ_{\N} = C_{\N} / κ_{\C\B} $, and $\ga =  κ_{\C\B} / κ_{\S\C} $ defined by
\begin{align}\label{eq:reparameterization}
\frac1{	ζ(ω) κ_{\C\B} } \!=\!  -i ω τ_C \!+\! 1 \!+\!\! \frac{ -i ω  \para{  τ_S   +  \frac{ τ_N }{ -i ω T\!1 + 1 } }    }{ -i ω \ga  \para{  τ_S   +  \frac{ τ_N }{ -i ω T\!1 + 1  } }  \!+\!1  } \,.
\end{align} 
The spectral decomposition of the thermal impedance in Eq.~(\ref{eq:theRm}), $ζ(ω) = ∑_{i=1,2,3} \, {A_i}/{[-iω + 1/m_i]}$,  has three characteristic times $m_{1,2,3}$. They are related to the parameters in the model via the set of equations, 
\begin{align}
				 m_1m_2m_3 =&   \ga τ_{\S} τ_{\C} \text{T1} \,,   \notag\\     
				m_1 +& m_2 + m_3 =  \text{T1} + τ_{\C} + (1 + \ga ) ( τ_{\N} + τ_{\S} )\,,  \notag\\ 
				 m_1 m_2 +& m_2 m_3 + m_3 m_1  \notag\\
				 	=&   \text{T1} τ_{\C} + ( 1 + \ga) τ_{\S} \text{T1}  + \ga τ_{\C} (τ_{\N} + τ_{\S} )  \,.	
\end{align}

\section{Fitting.}
\label{sec:fitting}

The fitting parameters are found by the gradient-descent minimization of the ``goodness'' function 
\begin{align}
&g(β(ω); \{λ_i\})\notag\\
&=∫\!\!dω\,\, β(ω)\brac{ζ(ω)^{\observed} - ζ(ω)_{\{λ_i\}}^{\model}}×\brac{\text{c.c.}}\,→\min 	
\end{align}
where $β(ω)$ is the weight function, and $λ_{i=1...6}$ are parameters in  Eq.~(\ref{eq:reparameterization}). $\brac{\text{c.c.}}$ is the complex conjugate of the first square bracket. The sensitivity of the fitting parameters (variances $Δλ_{i=1...6}$) to the changes in the measured $ζ(ω)^{\observed} $ is determined by the curvatures of the ``effective'' goodness function $G(β(ω); \{λ_i\})$, defined via 
\begin{align}
e^{ - G(β(ω); \{λ_i\}) } = \av{ e^{ - g(β(ω); \{λ_i\}) } } \,, 	
\end{align}
as
\begin{align}
Δλ_{i=1...6} = \sqrt{ [C^{-1} ]_{ii} }\,, \qquad \text{where} \qquad 
	C_{ij}  = \frac{d^2G(β(ω);  λ_{i}) }{ dλ_{i}dλ_{j} } 
\end{align}
Here $\av{⋯}$ is the average with respect to the noise in $ζ(ω)^{\observed}$ \cite{Hardybook}.  Assuming Gaussian noise with power spectrum $π(ω)$, the averaging $\av{⋯}$ over the noise is equivalent to setting the frequency-dependent ``temperature'' in the effective goodness function to the power spectrum $π(ω)$ or the weights $β(ω)$ to its inverse, $β(ω) \!=\! 1/ π(ω)$. The variances determined this way are shown as error bars in Figure 4. 

\section{Thermal susceptibility of a quasi-one-dimensional object}
\label{sec:xifunctions}

	The thermal susceptibility $ξ(sh,ω)$ describes the temperature change at the point $x=s$ at time $t$ in response to heat injection at $x=h$, at earlier time $t'$ (Figure 1),  
\begin{align}
		 -i ω C T(ω)_{x=s}  = \; ξ(sh,ω) \; q(ω)_{x=h} \,. 
\end{align}
	 It is determined by the heat flow dynamics in a quasi-one-dimensional (bar-like) object with boundary conditions of no heat influx at both ends, $x=0,L$, satisfied by the basis  
	 \begin{align}
	 φ_{k_n}(x) = \sqrt{2/L}\cos k_nx\,, \text{ where } k_n = πn/L \,, 	 \text{and } n=1...∞  	
	 \end{align}
	 is a positive integer.  One finds 
	\begin{align}
		ξ(sh;ω) \!=\! 1 \!+\!  2  ∑_{k_n} \! \frac{-iωτ}{ {-iωτ \!\!+\! (k_nL)^2 }} \!\cos k_n s \cos k_n h\,, 
	\end{align}
	where $τ  = L^2/D$ is the heat diffusion time across the bar, and $D$ is the heat diffusion coefficient, $D  = κ / c $. For $s,h=0,L$ at either end of the bar, $ξ(sh;ω)$ can be expressed in terms of elementary functions as, 
	\begin{align}
	ξ(00;ω ) =&     \sqrt{-iωτ}\coth\sqrt{-iωτ} \,, \notag\\
	ξ(0L;ω)^2 =&  ξ(00;ω)^2	+ iωτ \,. 		
	\end{align}

\section{Cross-relaxation}
\label{sec:crossrelaxation}

The thermodynamics of the nuclear spin subsystem can be described in a standard way \cite{Callen1985}, starting from the free energy, $dF_{\N} = - S_{\N} dT - M_{\N} dB$ (where $S_{\N}$ is its entropy and $M_{\N}$ is the magnetization) and then defining three thermodynamic coefficients, the heat capacity ($C_{\N}$), the magnetic susceptibility ($χ_{\N}$) and the nuclear magnetocaloric coefficient ($\mcal_{\N} $) via  
\begin{align}\label{eq:freenucl}
	\matr{ dS_{\N} \\ dM_{\N} } = 
\matr{ 
	 	\frac{C_{\N}}{T_{\N}} & \mcal_{\N} \\ 
	 	\mcal_{\N} & χ_{\N} }
	\matr{ dT_{\N} \\ dB } \,.
	\end{align} 
The heat exchange rate of nuclear spins is determined by the first line in this equation,  
$T_{\N} \frap{dS_{\N}}{dt} = C_{\N} \frap{dT_{\N} }{dt} + T_{\N} \mcal_{\N} \frap{dB}{dt}$. It equates the heat influx $q^{\NfromS} = T_{\N}\frap{dS_{\N}}{dt}$ from the lattice to the difference between the rate of change of the energy stored in the nuclear spin subsystem and the work done per unit time by an external magnetic field. For nuclear spins, the magnetocaloric coefficient \cite{Callen1985}  $\mcal_{\N} = \frap{dS_{\N}}{dB}_T =  \frap{dM_{\N}}{dT}_B$ is proportional to their magnetization, $\mcal_{\N}  = - \fra{M_{\N}}{T_{\N}} $, and, therefore, the magnetocaloric term in the energy balance,  $T_{\N} \mcal_{\N} \frap{dB}{dt}$ is equal to the work done per second by the external magnetic field on the nuclear spins, $-M\frap{dB}{dt}$. 

When several nuclear spin species (or several inequivalent lattice sites) are out of equilibrium, the energy exchange rates between nuclear spins and the lattice are determined not only by a distinct nuclear spin-lattice time for each isotope species but also by cross-relaxation coefficients \cite{Abragam1961, Hebel1963}. In the context of calorimetric measurements, the energy exchange between spins and lattice is determined by the symmetric matrix of relaxation coefficients \cite{Onsager1930, Onsager1931}, and therefore, the effective nuclear spin-lattice contact heat conductance is described by a symmetric matrix of transport coefficients. 

The entropy relaxation is governed by 
\begin{align}
	&\frac{d(S_{\N1}+S_{\N2}+S_{\S})}{dt} \notag\\
	&=
	  	q^{{\N1}\from{\S}} \para{ \frac1{T_{\N1}}-\frac1{T_{\S}}}  	
	+  q^{{\N2}\from{\S}} \para{ \frac1{T_{\N2}}-\frac1{T_{\S}}} \,.
\end{align} 
The heat exchange between two isotopic components and the electron-phonon subsystems is described by a symmetric 2x2 matrix $ Γ_{ij} $ of kinetic coefficients \cite{Onsager1930, Onsager1931}, $\fra{ dx_i}{dt}  =  ∑_{j=1,2}  Γ_{ij} X_j$ or, explicitly, 
%\begin{align} 
%	q^{{\N i}\from{\S}} = ∑_{j=1,2}  Γ_{ij}  \para{ \frac1{T_{\N j}}-\frac1{T_{\S}}} \,. 	
%\end{align}
\begin{align}
	\matr{ 
	q^{{\N 1}\from{\S}} \\
	q^{{\N 2}\from{\S}} } 
	= \matr{  
	Γ_{11} & Γ_{12} \\
	 Γ_{12} & Γ_{22} } 
	 \matr{  
	 \frac1{T_{\N 1}}-\frac1{T_{\S}} \\
	 \frac1{T_{\N 2}}-\frac1{T_{\S}} 
	 } \,. 	
\end{align}
The term proportional to $Γ_{12}$ describes cross-relaxation \cite{Bloembergen1959, Pershan1960, MacLaughlin1970, Abragam1961, Hebel1963}  in the two-isotope system. These equations replace Eq.~(\ref{eq:spinlattice}) for  multi-isotope nuclear spins.

\bibliographystyle{apsrev4-1}
%\bibliography{heatrefsark}

\begin{thebibliography}{40}%
\makeatletter
\providecommand \@ifxundefined [1]{%
 \@ifx{#1\undefined}
}%
\providecommand \@ifnum [1]{%
 \ifnum #1\expandafter \@firstoftwo
 \else \expandafter \@secondoftwo
 \fi
}%
\providecommand \@ifx [1]{%
 \ifx #1\expandafter \@firstoftwo
 \else \expandafter \@secondoftwo
 \fi
}%
\providecommand \natexlab [1]{#1}%
\providecommand \enquote  [1]{``#1''}%
\providecommand \bibnamefont  [1]{#1}%
\providecommand \bibfnamefont [1]{#1}%
\providecommand \citenamefont [1]{#1}%
\providecommand \href@noop [0]{\@secondoftwo}%
\providecommand \href [0]{\begingroup \@sanitize@url \@href}%
\providecommand \@href[1]{\@@startlink{#1}\@@href}%
\providecommand \@@href[1]{\endgroup#1\@@endlink}%
\providecommand \@sanitize@url [0]{\catcode `\\12\catcode `\$12\catcode `\&12\catcode `\#12\catcode `\^12\catcode `\_12\catcode `\%12\relax}%
\providecommand \@@startlink[1]{}%
\providecommand \@@endlink[0]{}%
\providecommand \url  [0]{\begingroup\@sanitize@url \@url }%
\providecommand \@url [1]{\endgroup\@href {#1}{\urlprefix }}%
\providecommand \urlprefix  [0]{URL }%
\providecommand \Eprint [0]{\href }%
\providecommand \doibase [0]{http://dx.doi.org/}%
\providecommand \selectlanguage [0]{\@gobble}%
\providecommand \bibinfo  [0]{\@secondoftwo}%
\providecommand \bibfield  [0]{\@secondoftwo}%
\providecommand \translation [1]{[#1]}%
\providecommand \BibitemOpen [0]{}%
\providecommand \bibitemStop [0]{}%
\providecommand \bibitemNoStop [0]{.\EOS\space}%
\providecommand \EOS [0]{\spacefactor3000\relax}%
\providecommand \BibitemShut  [1]{\csname bibitem#1\endcsname}%
\let\auto@bib@innerbib\@empty
%</preamble>
\bibitem [{\citenamefont {Tagliati}\ \emph {et~al.}(2012)\citenamefont {Tagliati}, \citenamefont {Krasnov},\ and\ \citenamefont {Rydh}}]{Tagliati2012}%
  \BibitemOpen
  \bibfield  {author} {\bibinfo {author} {\bibfnamefont {S.}~\bibnamefont {Tagliati}}, \bibinfo {author} {\bibfnamefont {V.~M.}\ \bibnamefont {Krasnov}}, \ and\ \bibinfo {author} {\bibfnamefont {A.}~\bibnamefont {Rydh}},\ }\href {\doibase 10.1063/1.4717676} {\bibfield  {journal} {\bibinfo  {journal} {Review of Scientific Instruments}\ }\textbf {\bibinfo {volume} {83}},\ \bibinfo {pages} {055107} (\bibinfo {year} {2012})}\BibitemShut {NoStop}%
\bibitem [{\citenamefont {Willa}\ \emph {et~al.}(2017)\citenamefont {Willa}, \citenamefont {Diao}, \citenamefont {Campanini}, \citenamefont {Welp}, \citenamefont {Divan}, \citenamefont {Hudl}, \citenamefont {Islam}, \citenamefont {Kwok},\ and\ \citenamefont {Rydh}}]{Willa2017}%
  \BibitemOpen
  \bibfield  {author} {\bibinfo {author} {\bibfnamefont {K.}~\bibnamefont {Willa}}, \bibinfo {author} {\bibfnamefont {Z.}~\bibnamefont {Diao}}, \bibinfo {author} {\bibfnamefont {D.}~\bibnamefont {Campanini}}, \bibinfo {author} {\bibfnamefont {U.}~\bibnamefont {Welp}}, \bibinfo {author} {\bibfnamefont {R.}~\bibnamefont {Divan}}, \bibinfo {author} {\bibfnamefont {M.}~\bibnamefont {Hudl}}, \bibinfo {author} {\bibfnamefont {Z.}~\bibnamefont {Islam}}, \bibinfo {author} {\bibfnamefont {W.-K.}\ \bibnamefont {Kwok}}, \ and\ \bibinfo {author} {\bibfnamefont {A.}~\bibnamefont {Rydh}},\ }\href {\doibase 10.1063/1.5016592} {\bibfield  {journal} {\bibinfo  {journal} {Review of Scientific Instruments}\ }\textbf {\bibinfo {volume} {88}},\ \bibinfo {pages} {125108} (\bibinfo {year} {2017})}\BibitemShut {NoStop}%
\bibitem [{\citenamefont {Purcell}\ \emph {et~al.}(1945)\citenamefont {Purcell}, \citenamefont {Torrey},\ and\ \citenamefont {Pound}}]{Purcell1945}%
  \BibitemOpen
  \bibfield  {author} {\bibinfo {author} {\bibfnamefont {E.~M.}\ \bibnamefont {Purcell}}, \bibinfo {author} {\bibfnamefont {H.~C.}\ \bibnamefont {Torrey}}, \ and\ \bibinfo {author} {\bibfnamefont {R.~V.}\ \bibnamefont {Pound}},\ }\href@noop {} {\bibfield  {journal} {\bibinfo  {journal} {Phys. Rev.}\ }\textbf {\bibinfo {volume} {69}},\ \bibinfo {pages} {37} (\bibinfo {year} {1945})}\BibitemShut {NoStop}%
\bibitem [{\citenamefont {Purcell}(1951)}]{Purcell1951}%
  \BibitemOpen
  \bibfield  {author} {\bibinfo {author} {\bibfnamefont {E.~M.}\ \bibnamefont {Purcell}},\ }\href@noop {} {\bibfield  {journal} {\bibinfo  {journal} {Physica}\ }\textbf {\bibinfo {volume} {17}},\ \bibinfo {pages} {282} (\bibinfo {year} {1951})}\BibitemShut {NoStop}%
\bibitem [{\citenamefont {Bloembergen}\ \emph {et~al.}(1948)\citenamefont {Bloembergen}, \citenamefont {Purcell},\ and\ \citenamefont {Pound}}]{Bloembergen1948}%
  \BibitemOpen
  \bibfield  {author} {\bibinfo {author} {\bibfnamefont {N.}~\bibnamefont {Bloembergen}}, \bibinfo {author} {\bibfnamefont {E.~M.}\ \bibnamefont {Purcell}}, \ and\ \bibinfo {author} {\bibfnamefont {R.~V.}\ \bibnamefont {Pound}},\ }\href@noop {} {\bibfield  {journal} {\bibinfo  {journal} {Physical Review}\ }\textbf {\bibinfo {volume} {73}},\ \bibinfo {pages} {679} (\bibinfo {year} {1948})}\BibitemShut {NoStop}%
\bibitem [{\citenamefont {Abragam}(1961)}]{Abragam1961}%
  \BibitemOpen
  \bibfield  {author} {\bibinfo {author} {\bibfnamefont {A.}~\bibnamefont {Abragam}},\ }\href@noop {} {\emph {\bibinfo {title} {The principles of nuclear magnetism}}}\ (\bibinfo  {publisher} {Oxford university press},\ \bibinfo {year} {1961})\BibitemShut {NoStop}%
\bibitem [{\citenamefont {Richardson}\ and\ \citenamefont {Smith}(2018)}]{Richardson2018}%
  \BibitemOpen
  \bibfield  {author} {\bibinfo {author} {\bibfnamefont {R.~C.}\ \bibnamefont {Richardson}}\ and\ \bibinfo {author} {\bibfnamefont {E.~N.}\ \bibnamefont {Smith}},\ }\href@noop {} {\emph {\bibinfo {title} {Experimental techniques in condensed matter physics at low temperatures}}}\ (\bibinfo  {publisher} {CRC Press},\ \bibinfo {year} {2018})\BibitemShut {NoStop}%
\bibitem [{\citenamefont {Pobell}(2007)}]{Pobell2007}%
  \BibitemOpen
  \bibfield  {author} {\bibinfo {author} {\bibfnamefont {F.}~\bibnamefont {Pobell}},\ }\href@noop {} {\emph {\bibinfo {title} {Matter and methods at low temperatures}}}\ (\bibinfo  {publisher} {Springer Berlin, Heidelberg},\ \bibinfo {year} {2007})\BibitemShut {NoStop}%
\bibitem [{\citenamefont {Sullivan}\ and\ \citenamefont {Seidel}(1968)}]{Sullivan1968}%
  \BibitemOpen
  \bibfield  {author} {\bibinfo {author} {\bibfnamefont {P.~F.}\ \bibnamefont {Sullivan}}\ and\ \bibinfo {author} {\bibfnamefont {G.}~\bibnamefont {Seidel}},\ }\href@noop {} {\bibfield  {journal} {\bibinfo  {journal} {Physical Review}\ }\textbf {\bibinfo {volume} {173}},\ \bibinfo {pages} {679} (\bibinfo {year} {1968})}\BibitemShut {NoStop}%
\bibitem [{\citenamefont {Wang}\ \emph {et~al.}(1988)\citenamefont {Wang}, \citenamefont {Campbell}, \citenamefont {Tsui},\ and\ \citenamefont {Cho}}]{Tsui1988}%
  \BibitemOpen
  \bibfield  {author} {\bibinfo {author} {\bibfnamefont {J.~K.}\ \bibnamefont {Wang}}, \bibinfo {author} {\bibfnamefont {J.~H.}\ \bibnamefont {Campbell}}, \bibinfo {author} {\bibfnamefont {D.~C.}\ \bibnamefont {Tsui}}, \ and\ \bibinfo {author} {\bibfnamefont {A.~Y.}\ \bibnamefont {Cho}},\ }\href@noop {} {\bibfield  {journal} {\bibinfo  {journal} {Physical Review B}\ }\textbf {\bibinfo {volume} {38}},\ \bibinfo {pages} {6174} (\bibinfo {year} {1988})}\BibitemShut {NoStop}%
\bibitem [{\citenamefont {Wang}\ \emph {et~al.}(1992)\citenamefont {Wang}, \citenamefont {Tsui}, \citenamefont {Santos},\ and\ \citenamefont {Shayegan}}]{Shayegan1991}%
  \BibitemOpen
  \bibfield  {author} {\bibinfo {author} {\bibfnamefont {J.~K.}\ \bibnamefont {Wang}}, \bibinfo {author} {\bibfnamefont {D.~C.}\ \bibnamefont {Tsui}}, \bibinfo {author} {\bibfnamefont {M.}~\bibnamefont {Santos}}, \ and\ \bibinfo {author} {\bibfnamefont {M.}~\bibnamefont {Shayegan}},\ }\href@noop {} {\bibfield  {journal} {\bibinfo  {journal} {Physical Review B}\ }\textbf {\bibinfo {volume} {45}},\ \bibinfo {pages} {4384} (\bibinfo {year} {1992})}\BibitemShut {NoStop}%
\bibitem [{\citenamefont {MacLaughlin}\ and\ \citenamefont {Butterworth}(1966)}]{Maclaughlin1966}%
  \BibitemOpen
  \bibfield  {author} {\bibinfo {author} {\bibfnamefont {D.~E.}\ \bibnamefont {MacLaughlin}}\ and\ \bibinfo {author} {\bibfnamefont {J.}~\bibnamefont {Butterworth}},\ }\href@noop {} {\bibfield  {journal} {\bibinfo  {journal} {Physics Letters}\ }\textbf {\bibinfo {volume} {23}},\ \bibinfo {pages} {291} (\bibinfo {year} {1966})}\BibitemShut {NoStop}%
\bibitem [{\citenamefont {MacLaughlin}\ \emph {et~al.}(1971)\citenamefont {MacLaughlin}, \citenamefont {Williamson},\ and\ \citenamefont {Butterworth}}]{Maclaughlin1971}%
  \BibitemOpen
  \bibfield  {author} {\bibinfo {author} {\bibfnamefont {D.~E.}\ \bibnamefont {MacLaughlin}}, \bibinfo {author} {\bibfnamefont {J.~D.}\ \bibnamefont {Williamson}}, \ and\ \bibinfo {author} {\bibfnamefont {J.}~\bibnamefont {Butterworth}},\ }\href@noop {} {\bibfield  {journal} {\bibinfo  {journal} {Physical Review B}\ }\textbf {\bibinfo {volume} {4}},\ \bibinfo {pages} {60} (\bibinfo {year} {1971})}\BibitemShut {NoStop}%
\bibitem [{\citenamefont {ONeal}\ and\ \citenamefont {Phillips}(1965)}]{oneal1964}%
  \BibitemOpen
  \bibfield  {author} {\bibinfo {author} {\bibfnamefont {H.~R.}\ \bibnamefont {ONeal}}\ and\ \bibinfo {author} {\bibfnamefont {N.~E.}\ \bibnamefont {Phillips}},\ }\href@noop {} {\bibfield  {journal} {\bibinfo  {journal} {Physical Review A}\ }\textbf {\bibinfo {volume} {137}},\ \bibinfo {pages} {748} (\bibinfo {year} {1965})}\BibitemShut {NoStop}%
\bibitem [{\citenamefont {Stone}(2014)}]{Stone2014}%
  \BibitemOpen
  \bibfield  {author} {\bibinfo {author} {\bibfnamefont {N.~J.}\ \bibnamefont {Stone}},\ }\href {https://www-nds.iaea.org/publications/indc/indc-nds-0658.pdf} {\enquote {\bibinfo {title} {table of nuclear magnetic dipole and electric quadrupole moments},}\ } (\bibinfo {year} {2014}),\ \Eprint {http://arxiv.org/abs/INDC(NDS)-0658} {Atomic Data and Nuclear Data Tables:INDC(NDS)-0658} \BibitemShut {NoStop}%
\bibitem [{\citenamefont {Butterworth}\ and\ \citenamefont {MacLaughlin}(1968)}]{Maclaughlin1968}%
  \BibitemOpen
  \bibfield  {author} {\bibinfo {author} {\bibfnamefont {J.}~\bibnamefont {Butterworth}}\ and\ \bibinfo {author} {\bibfnamefont {D.~E.}\ \bibnamefont {MacLaughlin}},\ }\href@noop {} {\bibfield  {journal} {\bibinfo  {journal} {Physical Review Letters}\ }\textbf {\bibinfo {volume} {20}},\ \bibinfo {pages} {265} (\bibinfo {year} {1968})}\BibitemShut {NoStop}%
\bibitem [{\citenamefont {Williamson}\ and\ \citenamefont {MacLaughlin}(1973)}]{Maclaughlin1973}%
  \BibitemOpen
  \bibfield  {author} {\bibinfo {author} {\bibfnamefont {J.~D.}\ \bibnamefont {Williamson}}\ and\ \bibinfo {author} {\bibfnamefont {D.~E.}\ \bibnamefont {MacLaughlin}},\ }\href@noop {} {\bibfield  {journal} {\bibinfo  {journal} {Physical Review B}\ }\textbf {\bibinfo {volume} {8}},\ \bibinfo {pages} {125} (\bibinfo {year} {1973})}\BibitemShut {NoStop}%
\bibitem [{\citenamefont {Adams}\ \emph {et~al.}(1966)\citenamefont {Adams}, \citenamefont {Berry},\ and\ \citenamefont {Hewitt}}]{Adams1966}%
  \BibitemOpen
  \bibfield  {author} {\bibinfo {author} {\bibfnamefont {J.~E.}\ \bibnamefont {Adams}}, \bibinfo {author} {\bibfnamefont {L.}~\bibnamefont {Berry}}, \ and\ \bibinfo {author} {\bibfnamefont {R.~R.}\ \bibnamefont {Hewitt}},\ }\href@noop {} {\bibfield  {journal} {\bibinfo  {journal} {Physical Review}\ }\textbf {\bibinfo {volume} {143}},\ \bibinfo {pages} {164} (\bibinfo {year} {1966})}\BibitemShut {NoStop}%
\bibitem [{\citenamefont {Hammond}\ \emph {et~al.}(1966)\citenamefont {Hammond}, \citenamefont {Wikner},\ and\ \citenamefont {Kelly}}]{Hammond1966}%
  \BibitemOpen
  \bibfield  {author} {\bibinfo {author} {\bibfnamefont {R.~H.}\ \bibnamefont {Hammond}}, \bibinfo {author} {\bibfnamefont {E.~G.}\ \bibnamefont {Wikner}}, \ and\ \bibinfo {author} {\bibfnamefont {G.~M.}\ \bibnamefont {Kelly}},\ }\href@noop {} {\bibfield  {journal} {\bibinfo  {journal} {Physical Review}\ }\textbf {\bibinfo {volume} {143}},\ \bibinfo {pages} {275} (\bibinfo {year} {1966})}\BibitemShut {NoStop}%
\bibitem [{\citenamefont {Casimir}\ and\ \citenamefont {du~Pre}(1938)}]{Casimir1938}%
  \BibitemOpen
  \bibfield  {author} {\bibinfo {author} {\bibfnamefont {H.}~\bibnamefont {Casimir}}\ and\ \bibinfo {author} {\bibfnamefont {F.}~\bibnamefont {du~Pre}},\ }\href {\doibase 10.1016/S0031-8914(38)80164-6} {\bibfield  {journal} {\bibinfo  {journal} {Physica}\ }\textbf {\bibinfo {volume} {5}},\ \bibinfo {pages} {507} (\bibinfo {year} {1938})}\BibitemShut {NoStop}%
\bibitem [{\citenamefont {Vleck}(1940)}]{VanVleck1940}%
  \BibitemOpen
  \bibfield  {author} {\bibinfo {author} {\bibfnamefont {J.~H.~V.}\ \bibnamefont {Vleck}},\ }\href {\doibase 10.1103/PhysRev.57.426} {\bibfield  {journal} {\bibinfo  {journal} {Phys. Rev.}\ }\textbf {\bibinfo {volume} {57}},\ \bibinfo {pages} {426} (\bibinfo {year} {1940})}\BibitemShut {NoStop}%
\bibitem [{\citenamefont {Hebel}(1963)}]{Hebel1963}%
  \BibitemOpen
  \bibfield  {author} {\bibinfo {author} {\bibfnamefont {L.~C.}\ \bibnamefont {Hebel}},\ }\href@noop {} {\bibfield  {journal} {\bibinfo  {journal} {Solid State Physics}\ }\textbf {\bibinfo {volume} {15}},\ \bibinfo {pages} {409} (\bibinfo {year} {1963})}\BibitemShut {NoStop}%
\bibitem [{\citenamefont {Bloembergen}\ and\ \citenamefont {Rowland}(1953)}]{Bloembergen1953}%
  \BibitemOpen
  \bibfield  {author} {\bibinfo {author} {\bibfnamefont {N.}~\bibnamefont {Bloembergen}}\ and\ \bibinfo {author} {\bibfnamefont {T.~J.}\ \bibnamefont {Rowland}},\ }\href@noop {} {\bibfield  {journal} {\bibinfo  {journal} {Acta metallurgica}\ }\textbf {\bibinfo {volume} {1}},\ \bibinfo {pages} {731} (\bibinfo {year} {1953})}\BibitemShut {NoStop}%
\bibitem [{\citenamefont {Bloembergen}\ \emph {et~al.}(1959)\citenamefont {Bloembergen}, \citenamefont {Shapiro}, \citenamefont {Pershan},\ and\ \citenamefont {Artman}}]{Bloembergen1959}%
  \BibitemOpen
  \bibfield  {author} {\bibinfo {author} {\bibfnamefont {N.}~\bibnamefont {Bloembergen}}, \bibinfo {author} {\bibfnamefont {S.}~\bibnamefont {Shapiro}}, \bibinfo {author} {\bibfnamefont {P.~S.}\ \bibnamefont {Pershan}}, \ and\ \bibinfo {author} {\bibfnamefont {J.~O.}\ \bibnamefont {Artman}},\ }\href@noop {} {\bibfield  {journal} {\bibinfo  {journal} {Physical Review}\ }\textbf {\bibinfo {volume} {114}},\ \bibinfo {pages} {445} (\bibinfo {year} {1959})}\BibitemShut {NoStop}%
\bibitem [{\citenamefont {Onsager}(1931{\natexlab{a}})}]{Onsager1930}%
  \BibitemOpen
  \bibfield  {author} {\bibinfo {author} {\bibfnamefont {L.}~\bibnamefont {Onsager}},\ }\href@noop {} {\bibfield  {journal} {\bibinfo  {journal} {Physical Review}\ }\textbf {\bibinfo {volume} {37}},\ \bibinfo {pages} {405} (\bibinfo {year} {1931}{\natexlab{a}})}\BibitemShut {NoStop}%
\bibitem [{\citenamefont {Onsager}(1931{\natexlab{b}})}]{Onsager1931}%
  \BibitemOpen
  \bibfield  {author} {\bibinfo {author} {\bibfnamefont {L.}~\bibnamefont {Onsager}},\ }\href@noop {} {\bibfield  {journal} {\bibinfo  {journal} {Physical Review}\ }\textbf {\bibinfo {volume} {38}},\ \bibinfo {pages} {2265} (\bibinfo {year} {1931}{\natexlab{b}})}\BibitemShut {NoStop}%
\bibitem [{\citenamefont {Leggett}(1975)}]{Leggett1975}%
  \BibitemOpen
  \bibfield  {author} {\bibinfo {author} {\bibfnamefont {A.~J.}\ \bibnamefont {Leggett}},\ }\href@noop {} {\bibfield  {journal} {\bibinfo  {journal} {Reviews of Modern Physics}\ }\textbf {\bibinfo {volume} {47}},\ \bibinfo {pages} {331} (\bibinfo {year} {1975})}\BibitemShut {NoStop}%
\bibitem [{\citenamefont {Korringa}(1950)}]{Korringa1950}%
  \BibitemOpen
  \bibfield  {author} {\bibinfo {author} {\bibfnamefont {J.}~\bibnamefont {Korringa}},\ }\href@noop {} {\bibfield  {journal} {\bibinfo  {journal} {Physica}\ }\textbf {\bibinfo {volume} {16}},\ \bibinfo {pages} {601} (\bibinfo {year} {1950})}\BibitemShut {NoStop}%
\bibitem [{\citenamefont {Landau}(1957)}]{Landau1956}%
  \BibitemOpen
  \bibfield  {author} {\bibinfo {author} {\bibfnamefont {L.~D.}\ \bibnamefont {Landau}},\ }\href@noop {} {\bibfield  {journal} {\bibinfo  {journal} {Sov. Phys. JETP}\ }\textbf {\bibinfo {volume} {3}},\ \bibinfo {pages} {920} (\bibinfo {year} {1957})}\BibitemShut {NoStop}%
\bibitem [{\citenamefont {Leggett}(1968)}]{Leggett1968}%
  \BibitemOpen
  \bibfield  {author} {\bibinfo {author} {\bibfnamefont {A.~J.}\ \bibnamefont {Leggett}},\ }\href@noop {} {\bibfield  {journal} {\bibinfo  {journal} {Annals of Physics}\ }\textbf {\bibinfo {volume} {46}},\ \bibinfo {pages} {76} (\bibinfo {year} {1968})}\BibitemShut {NoStop}%
\bibitem [{\citenamefont {Anderson}\ and\ \citenamefont {Redfield}(1959)}]{AndersonRedfield1959}%
  \BibitemOpen
  \bibfield  {author} {\bibinfo {author} {\bibfnamefont {A.~G.}\ \bibnamefont {Anderson}}\ and\ \bibinfo {author} {\bibfnamefont {A.~G.}\ \bibnamefont {Redfield}},\ }\href@noop {} {\bibfield  {journal} {\bibinfo  {journal} {Physical Review}\ }\textbf {\bibinfo {volume} {116}},\ \bibinfo {pages} {583} (\bibinfo {year} {1959})}\BibitemShut {NoStop}%
\bibitem [{\citenamefont {Moriya}(1963)}]{Morya1963}%
  \BibitemOpen
  \bibfield  {author} {\bibinfo {author} {\bibfnamefont {T.}~\bibnamefont {Moriya}},\ }\href@noop {} {\bibfield  {journal} {\bibinfo  {journal} {Journal of the Physical Society of Japan}\ }\textbf {\bibinfo {volume} {18}},\ \bibinfo {pages} {516} (\bibinfo {year} {1963})}\BibitemShut {NoStop}%
\bibitem [{\citenamefont {Landau}\ \emph {et~al.}(1981)\citenamefont {Landau}, \citenamefont {Lifschitz},\ and\ \citenamefont {Pitajewski}}]{LandauVol10}%
  \BibitemOpen
  \bibfield  {author} {\bibinfo {author} {\bibfnamefont {L.~D.}\ \bibnamefont {Landau}}, \bibinfo {author} {\bibfnamefont {E.~M.}\ \bibnamefont {Lifschitz}}, \ and\ \bibinfo {author} {\bibfnamefont {L.~P.}\ \bibnamefont {Pitajewski}},\ }\href@noop {} {\emph {\bibinfo {title} {Course of theoretical physics. vol. 10: Physical kinetics}}}\ (\bibinfo  {publisher} {Oxford},\ \bibinfo {year} {1981})\BibitemShut {NoStop}%
\bibitem [{\citenamefont {Tanuma}\ \emph {et~al.}(2011)\citenamefont {Tanuma}, \citenamefont {Powell},\ and\ \citenamefont {Penn}}]{Tanuma2011}%
  \BibitemOpen
  \bibfield  {author} {\bibinfo {author} {\bibfnamefont {S.}~\bibnamefont {Tanuma}}, \bibinfo {author} {\bibfnamefont {C.~J.}\ \bibnamefont {Powell}}, \ and\ \bibinfo {author} {\bibfnamefont {D.~R.}\ \bibnamefont {Penn}},\ }\href {\doibase 10.1002/sia.3522} {\bibfield  {journal} {\bibinfo  {journal} {Surface and Interface Analysis}\ }\textbf {\bibinfo {volume} {43}},\ \bibinfo {pages} {689} (\bibinfo {year} {2011})}\BibitemShut {NoStop}%
\bibitem [{\citenamefont {Chandrasekhar}(1960)}]{Chandrasekhar1960}%
  \BibitemOpen
  \bibfield  {author} {\bibinfo {author} {\bibfnamefont {S.}~\bibnamefont {Chandrasekhar}},\ }\href@noop {} {\emph {\bibinfo {title} {Radiative Transfer}}}\ (\bibinfo  {publisher} {Dover Publications},\ \bibinfo {year} {1960})\BibitemShut {NoStop}%
\bibitem [{\citenamefont {Lenk}(1979)}]{Lenk1979}%
  \BibitemOpen
  \bibfield  {author} {\bibinfo {author} {\bibfnamefont {R.}~\bibnamefont {Lenk}},\ }\href@noop {} {\bibfield  {journal} {\bibinfo  {journal} {Progress in Nuclear Magnetic Resonance Spectroscopy}\ }\textbf {\bibinfo {volume} {13}},\ \bibinfo {pages} {271} (\bibinfo {year} {1979})}\BibitemShut {NoStop}%
\bibitem [{\citenamefont {Hardy}\ and\ \citenamefont {Broun}()}]{Hardybook}%
  \BibitemOpen
  \bibfield  {author} {\bibinfo {author} {\bibfnamefont {W.}~\bibnamefont {Hardy}}\ and\ \bibinfo {author} {\bibfnamefont {D.}~\bibnamefont {Broun}},\ }\href@noop {} {\emph {\bibinfo {title} {Methods of Experimental Physics}}}\ (\bibinfo  {publisher} {Unpublished})\BibitemShut {NoStop}%
\bibitem [{\citenamefont {Callen}(1985)}]{Callen1985}%
  \BibitemOpen
  \bibfield  {author} {\bibinfo {author} {\bibfnamefont {H.~B.}\ \bibnamefont {Callen}},\ }\href@noop {} {\emph {\bibinfo {title} {Thermodynamics and an Introduction to Thermostatistics}}}\ (\bibinfo {year} {1985})\BibitemShut {NoStop}%
\bibitem [{\citenamefont {Pershan}(1960)}]{Pershan1960}%
  \BibitemOpen
  \bibfield  {author} {\bibinfo {author} {\bibfnamefont {P.~S.}\ \bibnamefont {Pershan}},\ }\href@noop {} {\bibfield  {journal} {\bibinfo  {journal} {Physical Review}\ }\textbf {\bibinfo {volume} {117}},\ \bibinfo {pages} {109} (\bibinfo {year} {1960})}\BibitemShut {NoStop}%
\bibitem [{\citenamefont {Mansfield}\ \emph {et~al.}(1970)\citenamefont {Mansfield}, \citenamefont {MacLaughlin},\ and\ \citenamefont {Butterworth}}]{MacLaughlin1970}%
  \BibitemOpen
  \bibfield  {author} {\bibinfo {author} {\bibfnamefont {P.}~\bibnamefont {Mansfield}}, \bibinfo {author} {\bibfnamefont {D.~E.}\ \bibnamefont {MacLaughlin}}, \ and\ \bibinfo {author} {\bibfnamefont {J.}~\bibnamefont {Butterworth}},\ }\href@noop {} {\bibfield  {journal} {\bibinfo  {journal} {J. Phys. C: Solid State Phys.}\ }\textbf {\bibinfo {volume} {3}},\ \bibinfo {pages} {1071} (\bibinfo {year} {1970})}\BibitemShut {NoStop}%
\end{thebibliography}

\end{document}